\documentclass[12pt]{article} \usepackage{amssymb}
\usepackage{amsmath} \usepackage{graphicx}
\usepackage{theorem,amssymb,amsbsy,latexsym}

\newcommand{\BZ}{\mathrm{BZ}}

\newcommand{\Z}{\mathbb{Z}} 
\newcommand{\N}{\mathbb{N}} 
\newcommand{\R}{\mathbb{R}} 

\newcommand{\nd}{{\phantom\dag}}

\newcommand{\vac}{\Omega}

\newcommand{\cN}{{\cal N}} 
\newcommand{\cE}{{\cal E}}  

\newcommand{\mbf}[1]{{\boldsymbol {#1} }}

\newcommand{\ii}{{\rm i}}
\newcommand{\dd}{{\rm d}}

\newcommand{\ve}{{\bf e}}
\newcommand{\vx}{{\bf x}}
\newcommand{\vy}{{\bf y}}
\newcommand{\vk}{{\bf k}}
\newcommand{\vn}{{\bf n}}
\newcommand{\vp}{{\bf p}}
\newcommand{\vQ}{{\bf Q}}
\newcommand{\vzero}{{\mbf 0}}

\newcommand{\Ref}[1]{(\ref{#1})}
\newcommand{\eps}{\epsilon}
\newcommand{\tPiL}{\mbox{$\frac{2\pi}{L}$}} 
\newcommand{\tL}{\mbox{$\frac{1}{L}$}} 
\newcommand{\half}{\mbox{$\frac{1}{2}$}}

\begin{document}
\begin{flushright}
April 14, 2010
\end{flushright}
\vspace{.4cm}

\begin{center}

{\Large \bf A two dimensional analogue of the Luttinger model}

\vspace{1 cm}

{\bf Edwin Langmann} \\[2mm] {\it Theoretical Physics, KTH\\ AlbaNova\\
SE-10691 Stockholm, Sweden}\\ {\tt langmann@kth.se} \\[5mm]

\end{center}

\begin{abstract}
  We present a fermion model that is, as we suggest, a natural 2D
  analogue of the Luttinger model. We derive this model as a partial
  continuum limit of a 2D spinless lattice fermion system with local
  interactions and away from half filling. In this derivation, we use
  certain approximations that we motivate by physical arguments. We
  also present mathematical results that allow an exact treatment of
  parts of the degrees of freedom of this model by bosonization, and
  we propose to treat the remaining degrees of freedom by mean field
  theory.

\medskip

MSC-class: 81Q80; 81T25; 81T27

\end{abstract}

\section{Introduction} 
\label{sec1}
Lattice fermion systems in {\em one} dimension can be studied
successfully by various different methods and are by-now well
understood. One powerful approach is to perform a particular continuum
limit leading to a low-energy effective model which can be solved
analytically. This limit amounts to linearizing the 1D tight-binding
band relation at the locations of the non-interacting Fermi surface
(consisting of two points) and then removing the ultra-violet (UV)
cutoff.  In the simplest case of spinless lattice fermions with
short-range interactions away from half-filling, one thus obtains the
Luttinger model \cite{Lutt} which can be solved exactly using
bosonization \cite{LM}; see \cite{T,Th,J} for closely related
pioneering work. It is worth stressing that exact solubility means a
lot in this case: not only the partition function but also all Green's
functions of the model can be computed by analytical methods; see
e.g.\ \cite{LLL} and references therein. This method can be
generalized to 1D Hubbard type systems and is the basis of a paradigm
for 1D interacting fermion systems \cite{Haldane}; see e.g.\
\cite{Tsvelik} for a textbook presentation.  We note that bosonization
is based on precise mathematical results; see e.g.\ \cite{CR} or
\cite{DS}.

In this paper, we propose a similar approach in two dimensions.
Starting from a 2D analogue of the spinless lattice fermion system
mentioned above, we propose a particular partial continuum limit that
makes this system amenable to an analytical, non-perturbative
treatment in a finite doping regime away from half filling. This limit
leads to a model that, as we suggest, is a natural 2D analogue of the
Luttinger model. Different from 1D, only parts of the fermion degrees
of freedom of this model can be bosonized and thus treated exactly. We
propose to treat the remaining degrees of freedom using mean field
theory.  We also argue that there is a finite doping regime away from
half filling at which these remaining degrees of freedom have an
energy gap and that in this regime an exactly solvable truncation of
the model can be used. Our approach is applicable beyond weak
coupling.  Previous work on bosonization in 2D
\cite{Luther1,Mattis,Rice,Hlubina,Anderson,Luther2,review,P} is
shortly discussed at the end of Section~\ref{sec7}.

Our derivation of this 2D analogue of the Luttinger model relies on
approximations that we justify by physical arguments; see
Section~\ref{sec5}. Our proposal that this model gives a low energy
effective description of 2D lattice fermions is therefore, from a
mathematical point of view, a conjecture. At any rate, this model is
mathematically well-defined and can be treated rigorously. To our
opinion, it also has a certain mathematical beauty and naturalness;
see \Ref{Hn1}--\Ref{Ha1}.  The details of our approach are quite
involved \cite{EL,dWL1,dWL2}, and the aim of this letter is to
concisely present the main ideas and results.

We define our notation and the lattice fermion system that we take as
starting point in Section~\ref{sec2}.  Section~\ref{sec3} gives a
summary of our results, including a self-contained definition of the
2D analogue of the Luttinger model.  In Section~\ref{sec4}, we discuss
physical arguments and experiments that motivate and guide our
approach. Our derivation of the model and its treatment by
bosonization are outlined in Sections~\ref{sec5} and \ref{sec6},
respectively. Section~\ref{sec7} contains final remarks.

\section{Two dimensional  $t$-$V$ model}
\label{sec2}

\subsection{Notation}
\label{sec2.1}
We consider a square lattice with $\cN \gg 1$ sites and denote fermion
momenta as $\vk,\vk'$.  We write $\vk=(k_1,k_2)=k_+\ve_++k_-\ve_-$
with $\ve_\pm=(1,\pm 1)/\sqrt{2}$, i.e.\ $k_\pm = (k_1\pm
k_2)/\sqrt{2}$, and similarly for other 2D vectors $\vp$ etc.  We
introduce a lattice constant $a>0$ so that $-\pi/a\leq k_{1,2}
<\pi/a$, and our large distance cut-off $L$ (system size) is such that
$k_{\pm}\in(2\pi/L)(\Z+1/2)$ and $L/a\in 4\sqrt{2}(\N+1/2)$, i.e.\
$\cN=(L/a)^2$. The set of all such $\vk$ is denoted as $\BZ$
(Brillouin zone). We use the symbol $\vp$ for differences of fermion
momenta, and $\tilde\Lambda^*$ is the set of all $\vp$ such that
$p_{\pm}\in(2\pi/L)\Z$.  Fermion operators $\hat\psi(\vk)$ are defined
for $\vk\in\BZ$ and normalized such that
$\{\hat\psi(\vk),\hat\psi^\dag(\vk')\}=\delta_{\vk,\vk'}[L/(2\pi)]^2$,
and we use the same normalization for other fermion operators
$\hat\psi_{r,s}(\vk)$ introduced in Section~\ref{sec3}.  We write
$\hat\psi^\dag\hat\psi(\vk)$ short for
$\hat\psi^\dag(\vk)\hat\psi(\vk)$ etc. We use the abbreviations $
\tilde{a}=2\sqrt2 a $.  The symbol $\delta_a$ denotes the lattice
periodic Kronecker delta, i.e.\ $\delta_a(\vp)=1$ if $\vp\in
(2\pi/a)\Z^2$ and $0$ otherwise.  Finally, $[\vk+\vk']$ is the sum of
two momenta in $\BZ$ modulo $(2\pi/a)\Z^2$, i.e.\ it is equal to
$\vk+\vk'+(2\pi/a)\vn\in\BZ$ with appropriate $\vn\in\Z^2$.

\subsection{Definitions} 
\label{sec2.2}
The lattice fermion model we consider is the so-called 2D $t$-$V$
model. It describes spinless fermions on a 2D square lattice with
$\cN$ sites and hopping and repulsive density-density interactions
between nearest neighbor sites. It is defined by the Hamiltonian
\begin{equation} 
H_{tV}=H_0-\mu N + H_{\rm int}
\end{equation}
with the free part
\begin{equation} 
H_0 = \left( \tPiL \right)^2 \sum_{\vk\in\BZ} \eps(\vk)\,
\hat\psi^\dag \hat\psi(\vk)
\end{equation}
and $\eps(\vk) = -2t[\cos(ak_1) + \cos(ak_2)]$ the tight binding band
relation,
\begin{equation} 
N = \left( \tPiL \right)^2
\sum_{\vk\in\BZ} \hat\psi^\dag \hat\psi(\vk)
\end{equation}
the fermion number operator, and $\mu\in\R$ the chemical
potential. The interaction is
\begin{equation} 
\label{Hint} 
  H_{\rm int} =  \left( \tPiL \right)^6\sum_{\vk_j\in\BZ} 
  \hat\psi^\dag(\vk_1) \hat\psi(\vk_2) 
  \hat\psi^\dag(\vk_3) \hat\psi(\vk_4) 
  \hat u(\vk_1-\vk_2) \delta_a(\vk_1-\vk_2+\vk_3-\vk_4)  
\end{equation}
with $\hat u(\vp)= a^2 V [\cos(ap_1)+\cos(ap_2)]/(8\pi^2)$ the Fourier
transform of a nearest neighbor interaction.  The model parameters $t$
(hopping constant) and $V$ (coupling strength) both are positive.  The
{\em filling parameter} is defined as 
\begin{equation}
\label{nu}
\nu = \frac{\langle N\rangle}{\cN}
\end{equation} 
where $\langle \cdot \rangle$ denotes the ground state expectation
value. It is in the range $0\leq\nu\leq 1$ with half filling
corresponding to $\nu=1/2$, and $\nu-1/2$ is referred to as doping.

The Hamiltonian $H_{tV}$ is invariant under the particle-hole
transformation
\begin{equation} 
\label{PH}
\hat\psi(\vk)\to
\hat\psi^\dag(\vk+\vQ/a), \quad \mu\to V-\mu, \quad \nu\to 1-\nu
\end{equation}
for $\vQ=(\pi,\pi)$, up to an irrelevant additive constant. We
therefore can restrict ourselves to $\nu\geq 1/2$.

\section{Summary of results} 
\label{sec3}
The 2D analogue of the Luttinger model describes six flavors of
fermions $\hat\psi_{r,s}(\vk)$ with flavor indices $r=\pm$ and
$s=0,\pm$ and momenta $\vk$ in different Fourier spaces $\Lambda^*_s$
as follows,
\begin{equation}
\label{Lrs}
\begin{split}
  \Lambda_0^* =& \left\{ \vk \, |\; k_\pm \in \tPiL(\Z+\half),\;\; 
    -\frac{\pi}{\tilde{a}}\leq
    k_\pm < \frac{\pi}{\tilde{a}}\right\} \\
  \Lambda_{s=\pm}^* =& \left\{ \vk\, |\; k_\pm \in \tPiL(\Z+\half),\;\; 
    -\infty < k_s <
    \infty,\;\; -\frac{\pi}{\tilde{a}}\leq k_{-s} <
    \frac{\pi}{\tilde{a}}  \right\}. 
\end{split} 
\end{equation} 
We refer to the fermions with $s=0$ and $s=\pm$ as {\em nodal} and
{\em antinodal}, respectively. The Hamiltonian defining this model is
\begin{equation}
\label{HL}
H = H_n + H_a 
\end{equation} 
with the nodal part
\begin{equation}
\begin{split}
\label{Hn} 
  H_n = & \sum_{s=\pm}\Bigl( \left( \tPiL \right)^2
  \sum_{\vk\in\Lambda^*_{s}}\sum_{r=\pm} rv_F k_s :\!
  \hat\psi^\dag_{r,s}\hat\psi^{\nd}_{r,s}(\vk)\!: + \\ & \left( \tL
  \right)^2\sum_{\vp\in\tilde\Lambda^*}\chi(\vp)\bigl[
  g_1\sum_{r=\pm}\hat{J}^\dag_{r,s}\hat{J}^\nd_{-r,s}(\vp) +
  g_2\sum_{r,r'=\pm}\hat{J}^\dag_{r,s}\hat{J}^\nd_{r',-s}(\vp) \bigr]
  \Bigr)
\end{split} 
\end{equation} 
and the antinodal part (including nodal-antinodal interactions)
\begin{equation}
\begin{split}
\label{Ha}
  H_a = & \left( \tPiL \right)^2 \sum_{\vk\in\Lambda^*_0}\sum_{r=\pm}
  (-rc_Fk_+k_--\mu_a) :\!
  \hat\psi^\dag_{r,0}\hat\psi^{\nd}_{r,0}(\vk)\!: + \\ & \left( \tL
  \right)^2\sum_{\vp\in\tilde\Lambda^*}\bigl[
  g_3\sum_{r=\pm}\hat{J}^\dag_{r,0}\hat{J}^\nd_{-r,0}(\vp) +
  g_4\sum_{r,r',s=\pm}\chi(\vp)\hat{J}^\dag_{r,0}\hat{J}^\nd_{r',s}(\vp)
  \bigr]
\end{split} 
\end{equation} 
(the model parameters $v_F$, $c_F$, $g_j$, and $\mu_a$ are specified
further below).  The $\hat{J}_{r,s}$ are Fourier transformed and
normal ordered fermion densities defined as follows,
\begin{equation}
\begin{split}
  \label{Jrs}
  \hat{J}_{r,0}(\vp) =& \left( \tPiL \right)^2
  \sum_{\vk\in\Lambda^*_0}
  :\! \hat\psi^\dag_{r,0}(\vk-\vp)\hat\psi^{\nd}_{r,0}(\vk)\!: \\
  \hat{J}_{r,s=\pm}(\vp) =& \left( \tPiL \right)^2
  \sum_{\vk\in\Lambda^*_s} \sum_{n\in\Z} :\!
  \hat\psi^\dag_{r,s}(\vk-\vp)\hat\psi^{\nd}_{r,s}(\vk+2\pi
  n\ve_{-s}/\tilde{a})\!:
\end{split}
\end{equation} 
and $\hat{J}^\dag_{r,s}\hat{J}^\nd_{r',s'}(\vp)$ is short for
$\hat{J}_{r,s}(-\vp)\hat{J}_{r',s'}(\vp)$. We also use the cutoff
functions
\begin{equation}
\label{chi} 
\begin{split}
  \mbox{$\chi(\vp)=1$ if $-\pi/\tilde a \leq p_\pm \leq \pi/\tilde a$ and
    $0$ otherwise}.
\end{split} 
\end{equation}
The colons indicate normal ordering with respect to a non-interacting
groundstate $\vac$ defined by the conditions
\begin{equation} 
\label{vac}
\hat\psi^\nd_{r,\pm}(\vk)\vac = \hat\psi^\dag_{r,\pm}(-\vk)\vac = 0 \;
\mbox{ if $r k_\pm >0$}
\end{equation}
and $\hat\psi^\dag_{r,0}(\vk)\vac=0$ or
$\hat\psi^\nd_{r,0}(\vk)\vac=0$ if $r k_+k_->0$ or $<0$, respectively.
In the theorem below we set $\hat{J}_{r,s=\pm}(\vp)=0$ for
$|p_{-s}|> \pi/\tilde{a}$.

We first outline how the model above can be derived from the 2D
$t$-$V$ model using certain approximations (Section~\ref{sec5}). This
allows us to determine the model parameters as follows,
\begin{equation}
\label{vF} 
c_F= 2ta^2,\quad v_F = 2\sqrt 2 t a \sin(Q), \quad Q=\pi\nu 
\end{equation}
\begin{equation}
\label{gj}
g_1=2g_2=2V\sin^2(Q) a^2,\quad g_3=g_4=2V a^2
\end{equation}
\begin{equation} 
\label{mua}
\mu_a = -\left(4t
+\frac{V}4\right)\cos(Q)+V\cos^2(Q)\left(1-\frac{2Q}\pi\right)
\end{equation}
with $0<|\nu-1/2|<1/4$. The key result to bosonize the model above is
the following.

\medskip

\noindent \textbf{Theorem:} {\it (a) The nodal density operators
  obey the relations
  \begin{equation} 
\label{JJ} 
[\hat{J}_{r,s}(\vp),\hat{J}_{r',s'}(\vp')] =
    \delta_{r,r'}\delta_{s,s'} \frac{2\pi p_s}{\tilde{a}}
    \delta_{\vp,-\vp'} \left(\mbox{$\frac{L}{2\pi}$} \right)^2 \quad
    \forall r,s=\pm
\end{equation} 
and $\hat{J}_{r,s}(\vp)^\dag= \hat{J}_{r,s}(-\vp)$. Moreover,
$\hat{J}_{r,s}(\vp)\vac = 0$ $\forall \vp$ such that $rp_s\geq 0$.

\noindent (b) The nodal Hamiltonian in \Ref{Hn} is identical with
\begin{equation}
\label{Hnboson} 
  H_n = \left( \tL \right)^2\sum_{\vp\in\tilde\Lambda^*}
  \sum_{r,s=\pm}\Bigl( \pi \tilde{a}v_F
:\!\hat{J}_{r,s}\hat{J}_{r,s}(\vp)\!:+ 
  \chi(\vp)\bigl[
  g_1 \hat{J}_{r,s}\hat{J}_{-r,s}(\vp) +
  g_2\sum_{r'=\pm}\hat{J}_{r,s}\hat{J}_{r',-s}(\vp) \bigr] \Bigr) 
\end{equation} 
and well-defined provided that $V< 4\pi t/\sin(Q)$.}

\medskip

(Proof outlined in Section~\ref{sec6}.)

As discussed in Section~\ref{sec6}, his theorem provides the means to
exactly map $H_n$ to a bosons Hamiltonian.  Thus, the model above is
equivalent to a model of non-interacting bosons coupled linearly to
the antinodal fermions. It is therefore possible to integrate out the
nodal fermions exactly and thus obtain an effective antinodal model.
We also discuss how these results can be used to obtain physical
information about the 2D $t$-$V$ model.

\section{Physics motivation}
\label{sec4}
Our aim is to rewrite and modify the 2D $t$-$V$ model such that, (i)
the resulting model can be treated by bosonization, (ii) the low
energy physics is changed as little as possible.  For that we adopt
the following hypothesis that has been successful in 1D:

\noindent {\bf H1:} {\it There exists some underlying Fermi surface
  dominating the low energy physics, and we can modify, ignore or add
  degrees of freedom far away from this surface} (in the latter two
cases we need to correct the definition of doping, of course).

We note that, at half filling and without interactions, the relevant
Fermi surface is the square $|k_1\pm k_2|=\pi/a$ (the large diamond in
Figure~1).  We assume that the underlying Fermi surface remains close
to this square even in the presence of interactions and away from half
filling. We observe that the band relation has a qualitatively
different behavior in different regions of the Brillouin zone close to
such a surface.  To make this explicit, we select six representative
points $\vQ_{r,s}/a$ labeled by two indices $r=\pm$ and $s=0,\pm$ as
follows,
\begin{equation}
\label{vQ}
\begin{split}
 \vQ_{+,0} = (\pi,0), \quad \vQ_{-,0} = (0,\pi)\\ \vQ_{r,s} = (r
Q, r s Q) \; \mbox{ for $r,s=\pm$} 
\end{split}
\end{equation}
for some $Q\approx \pi/2$ (the dots in Figure~1). We then approximate
the band relation close to these points by truncating the Taylor
series at the lowest non-trivial order, $\eps(\vQ_{r,s}/a+\vk)\approx
\eps_{r,s}(\vk)$ with
\begin{equation} 
\label{epsrs} 
\eps_{r,0}(\vk) = -r c_F k_+k_- ,\quad \eps_{r, \pm}(\vk) = -4t\cos(Q) + r v_F
k_{\pm}
\end{equation}
and the constants $c_F$ and $v_F$ in \Ref{vF}.  At half-filling we
expect $Q=\pi/2$, and, for symmetry reasons, the underlying Fermi
surface should always contain four such points $\vQ_{r,\pm}$, $r=\pm$,
for some value of $Q$.  Thus, the band relation is hyperbolic close to
the vertices $\vQ_{r,0}/a$ of the square surface, and it is linear
with a constant Fermi velocity $v_F$ close to the midpoints of the
sides $\vQ_{r,\pm}/a$.  We use the terminology of experimental
physicists studying cuprate superconductors with ARPES \cite{ARPES}
and refer to the regions in Fourier space close to $\vQ_{r,0}/a$ and
$\vQ_{r,\pm}/a$ as antinodal and nodal, respectively.

As will become clear in the next section, Hypothesis~H1 is not enough
to justify the approximations we make. For that the following somewhat
stronger hypothesis is needed.

\noindent {\bf H1':} {\it The low energy properties of the model are
  not (much) changed if we modify the band structure and the
  interactions for fermion degrees of freedom far away from the six
  Fermi surface points $\vQ_{r,s}/a$ in \Ref{vQ}.}

There are physical arguments suggesting that, at half filling and
sufficiently large coupling, the 2D $t$-$V$ model has a charge density
wave (CDW) groundstate which is insulating; see e.g.\ Section~IV.A in
Ref.\ \cite{Shankar}. In mean fields theory this state is
characterized by a CDW gap $\Delta>0$ changing the band relations to
$\pm\sqrt{\eps(\vk)^2+\Delta^2}$, and this corresponds to a fully
gapped underlying Fermi surface.  Moreover, ARPES results on cuprate
superconductors show that there exists an interesting doping regime
away from half filling where, in these materials, the antinodal
fermions are gapped while the nodal fermions have no gap but Fermi
surface arcs \cite{ARPES}.  This suggests that the nodal- and
antinodal fermion degrees of freedom can have different physical
behavior, and it motivates us to rewrite the 2D $t$-$V$ model so that
nodal- and antinodal fermion degrees of freedom can be easier treated
by different computation methods. 

The model thus obtained can be further simplified by the following
hypothesis motivated by the ARPES results mentioned above.

\noindent {\bf H2:} {\it There exists a finite doping region away
  from half filling where the nodal points on the underlying Fermi
  surface move to $\vQ_{r,\pm}/a$ with $Q>\pi/2$, and in this regime
  the CDW gap is absent in the nodal regions while it is still present
  in the antinodal regions.}

In this regime, the parameter $Q$ is determined by doping. This
hypothesis is also suggested to us by results from renormalization
group studies \cite{FRS,HSFR} and mean field theory \cite{Schulz,LW}.

\section{Derivation of the model}
\label{sec5}
We now outline a derivation of the model in \Ref{HL}--\Ref{mua} from
the 2D $t$-$V$ model, emphasizing the approximations that are made.

\begin{figure}[ht]
\label{fig1}
\resizebox{!}{12cm}{\includegraphics{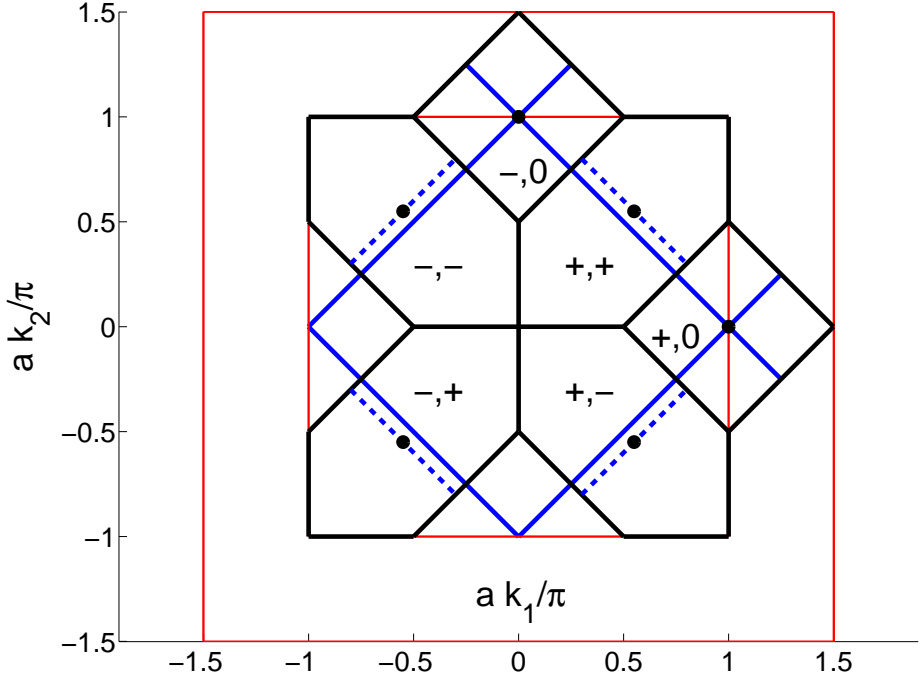}}
\caption{Division of the Brillouin zone in six regions
  $\Lambda^*_{r,s}$, $r=\pm$, $s=0,\pm$. The six dots mark the points
  $\vQ_{r,s}/a$ in \Ref{vQ} for filling $\nu=0.55$. The regions
  $\Lambda^*_{r,s}$ are equal to the six-sided polygons (for $s=\pm$)
  and small diamonds (for $s=0$) shifted such that the corresponding
  points $\vQ_{r,s}/a$ coincide with the origin of the coordinate
  system. Shown are also the half-filled square surface (large
  diamond) and the assumed interacting Fermi surface for the nodal
  fermions (dashed lines).}
\end{figure}

We divide the Brillouin zone in six regions as shown in Figure~1 (the
areas enclosed by boldface lines), and we label the fermion degrees of
freedom in these regions as follows,
\begin{equation} 
\label{hatpsi} 
\hat\psi_{r,s}(\vk) = \hat\psi([\vQ_{r,s}/a+\vk])
\end{equation}
where the momenta $\vk$ are now restricted to the small regions
$\Lambda^*_{r,s}$ containing only momenta satisfying $-\pi/\tilde
a\leq k_\pm<\pi/\tilde a$ for $s=0$, and $-\pi/\tilde a\leq k_{-s}<
\pi/\tilde a$, $-\pi/(2a)\leq k_1+r(Q-\pi/2)/a < \pi/(2a)$,
$-\pi/(2a)\leq k_2+rs(Q-\pi/2)/a <\pi/(2a)$ for $s=\pm$; see Figure~1.
We also assume $\pi/4<Q<3\pi/4$ and $Q\in \sqrt{2}\pi(a/L)\N$ (the
latter restriction becomes irrelevant in the limit
$L/a\to\infty$). We thus trade different regions in the Brillouin zone
for six different fermions flavors $r,s$. Note that
$\Lambda^*_{\pm,0}=\Lambda^*_0$ in \Ref{Lrs}, and the regions are
defined such that
\begin{equation}
\label{id} 
\sum_{\vk\in\BZ}f(\vk) =\sum_{r,s}\sum_{\vk\in\Lambda^*_{r,s}}
f([\vQ_{r,s}/a+\vk])
\end{equation}
for any function $f$ on $\BZ$ where, here and in the following,
sums over $r$ and $s$ are over $r=\pm$ and $s=0,\pm$ unless stated
otherwise. In particular, $N = \sum_{r,s} N_{r,s}$ with
\begin{equation} 
N_{r,s}=\left( \tPiL \right)^2
\sum_{\vk\in\Lambda^*_{r,s}} \hat\psi^\dag_{r,s}
\hat\psi^\nd_{r,s}(\vk)
\end{equation}
and $H_0=\sum_{r,s}(2\pi/L)^2\sum_{\vk\in\Lambda^*_{r,s}}
\eps(\vQ_{r,s}/a+\vk)\hat\psi^\dag_{r,s} \hat\psi^\nd_{r,s}(\vk)$.  We
now make our first approximation:

\noindent {\bf A1:} {\it Taylor expand the exact band relations in the
  vicinity of the Fermi surface points $\vQ_{r,s}/a$ and keep only the
  leading non-trivial terms, i.e.\ replace $\eps(\vQ_{r,s}/a+\vk)$ in
  $H_0$ above by $\eps_{r,s}(\vk)$ in \Ref{epsrs}.}

We thus obtain
\begin{equation} 
\label{HL0}
H_0 = \sum_{r,s} \left(\tPiL\right)^2
\sum_{\vk\in\Lambda^*_{r,s}} \eps_{r,s}(\vk)\,
\hat\psi_{r,s}^\dag \hat\psi^\nd_{r,s} (\vk). 
\end{equation}

In this approximation, we neglect terms of higher order in $a\vk$. We
expect that the low energy properties of the model are dominated by
states close to the Fermi surface points for which these higher order
terms are negligible.

To modify the interaction part of the Hamiltonian in a way suitable
for our purposes we use \Ref{id} to rewrite
\begin{equation}
\label{Hint1} 
\begin{split}  
  H_{\rm int} = \left( \tPiL
  \right)^6\sum_{r,s}\sum_{\vk_j\in\Lambda^*_{r,s}}
  \hat\psi^\dag_{r_1,s_1}(\vk_1) \hat\psi^{\nd}_{r_2,s_2}(\vk_2)
  \hat\psi^\dag_{r_3,s_3}(\vk_3) \hat\psi^{\nd}_{r_4,s_4}(\vk_4)
  \hat u(\vQ_{r_1,s_1}-\vQ_{r_2,s_2}+\vk_1-\vk_2)\\
  \times \delta_a(\vQ_{r_1,s_1}-\vQ_{r_2,s_2}+\vQ_{r_3,s_3} -
  \vQ_{r_4,s_4}+\vk_1-\vk_2+\vk_3-\vk_4)
\end{split} 
\end{equation} 
and make our second approximation:

\noindent {\bf A2:} {\it Replace the interaction Hamiltonian
  in \Ref{Hint1} by
\begin{equation}
\label{Hint2} 
\begin{split}  
  H_{\rm int} = \left( \tPiL
  \right)^6\sum_{r,s}\sum_{\vk_j\in\Lambda^*_{r,s}}
  \hat\psi^\dag_{r_1,s_1}(\vk_1) \hat\psi^{\nd}_{r_2,s_2}(\vk_2)
  \hat\psi^\dag_{r_3,s_3}(\vk_3) \hat\psi^{\nd}_{r_4,s_4}(\vk_4)
  \hat u(\vQ_{r_1,s_1}-\vQ_{r_2,s_2})\\
  \times \delta_a(\vQ_{r_1,s_1}-\vQ_{r_2,s_2}+\vQ_{r_3,s_3} -
  \vQ_{r_4,s_4})\delta_{\vk_1-\vk_2+\vk_3-\vk_4,\vzero}.
\end{split} 
\end{equation} 
}

Note that $\vQ_{r_1,s_1}-\vQ_{r_2,s_2}+\vQ_{r_3,s_3} - \vQ_{r_4,s_4}$
is either $\vzero$ or a ``large'' vector whose length is of order
$\pi/a$. We expect that the most important processes for the low
energy properties of the model are those where the $\vk_j$ are so
small that the Kronecker delta in \Ref{Hint1} can be non-zero only if
$\vk_1-\vk_2+\vk_3-\vk_4=\vzero$, and these processes are not much
affected by A2.

To simplify \Ref{Hint2} we have to find all solutions of
$\delta_a(\vQ_{r_1,s_1}-\vQ_{r_2,s_2}+\vQ_{r_3,s_3} -
\vQ_{r_4,s_4})=1$. We assume $Q\neq \pi/2$ and obtain after
straightforward computations \cite{EL}
\begin{equation}
\begin{split} 
\label{Hint3} 
  H_{\rm int} = \left( \tL
  \right)^2\sum_{\vp\in\tilde\Lambda^*}\bigl[
  &g_3\sum_{r=\pm}\hat{\rho}^\dag_{r,0}\hat{\rho}^\nd_{-r,0}(\vp) +
  g_1\sum_{r,s=\pm}\hat{\rho}^\dag_{r,s}\hat{\rho}^\nd_{-r,s}(\vp) \\+&
  g_2\sum_{r,r',s=\pm}\hat{\rho}^\dag_{r,s}\hat{\rho}^\nd_{r',-s}(\vp)+
  g_4\sum_{r,r',s=\pm}\hat{\rho}^\dag_{r,0}\hat{\rho}^\nd_{r',s}(\vp) \bigr]
\end{split} 
\end{equation} 
with
\begin{equation}
\label{rhors}
\hat{\rho}_{r,s}(\vp)=  \left( \tPiL \right)^2
\sum_{\vk_j\in\Lambda^*_{r,s}}
\hat\psi^\dag_{r,s}(\vk_1)\hat\psi^{\nd}_{r,s}(\vk_2)
\delta_{\vk_1+\vp,\vk_2}
\end{equation} 
and the coupling parameters in \Ref{gj}. 

The operators in \Ref{rhors} have the natural physical interpretation
as Fourier transformed density operators.  It is remarkable that we
only get interaction terms of density-density form.  Our assumption
$Q\neq \pi/2$ is important since otherwise one obtains additional
terms that cannot be treated in a simple manner by bosonization,
similarly as in 1D \cite{Tsvelik}. Such terms might cause a nodal
fermion gap.  This is consistent with our Hypothesis~H2 in
Section~\ref{sec4} since $Q=\pi/2$ corresponds to half filling, as
shown below.

To obtain a model that can be bosonized we want to remove the UV
cutoff for the nodal fermions orthogonal to the Fermi surface
arcs. Before we can do that we have to normal order all fermion
operators with respect to some suitable reference state $\vac$
(``Dirac sea'') with fixed fermion density $\nu\neq 1/2$. We choose
this state such that the nodal fermion states are filled up to the
Fermi surface through the points $\vQ_{r,\pm}/a$ parallel to the
half-filled one (the four dashed lines in Figure~1). For simplicity we
also assume that the antinodal fermions are half-filled (one can show
that this follows from our Hypothesis~H2), but this is not essential
(see \cite{dWL1} for a more general treatment). This reference state
can be defined by the conditions in \Ref{vac} {\em ff}.  By simple
geometric considerations we find that the contributions of the
different regions $r,s$ to the total filling $\nu$ of this state are
$\nu_{r,0} =1/16$ and $\nu_{r,s=\pm}=( Q/\pi -1/8)/4$, respectively,
and thus $\nu = \sum_{r,s}\nu_{r,s} = Q/\pi$. 

Normal ordering of the Hamiltonian $H_0-\mu N+H_{\rm int}$ amounts to
the following: rewrite $H_{\rm int}$ in terms of the normal ordered
fermion densities
\begin{equation} 
  \hat{J}_{r,s}(\vp) = \; :\!\hat{\rho}_{r,s}(\vp)\!: 
\end{equation}
and normal order $H_0$ and $N$, dropping an irrelevant additive
constant. An important consequence of this and Approximation~A1 is a
change of the chemical potential term $-\mu N$ in the Hamiltonian to
$-\sum_{r,s}\mu_{r,s}:\!N_{r,s}\!:$, and we find by straightforward
computations the following renormalized chemical potentials:
$\mu_{r,0}=\mu-(2Q/\pi)V\equiv\mu_a$ and $\mu_{r,s=\pm}= \mu
+4t\cos(Q)- [2Q\sin^2(Q)/\pi +\cos^2(Q)+\cos(Q)/4]V\equiv\mu_n$.  To
have an underlying Fermi surface as assumed in H2 we set
$\mu_n=0$. This fixes $\mu$ and implies \Ref{mua}. Note that $\mu_a$
changes sign under the transformation $Q\to \pi-Q$, as required by
invariance under particle-hole transformations.

By assumption the groundstate expectation values of $:\!N_{r,s}\!:$
all are zero in the parameter regime of interest to us, and thus the
filling factor of the groundstate is identical with $\nu=Q/\pi$. This
fixes $Q$ as in \Ref{vF}.

We now can partially remove the UV cutoff for the nodal fermions. We
do this by a sequence of three approximations:

\noindent {\bf A3:} {\it Replace in the interaction part of the
  Hamiltonian the normal ordered nodal densities
  $\hat{J}_{r,s=\pm}(\vp)$ by $\chi(\vp)\hat{J}_{r,s=\pm}(\vp)$ with
  the cutoff functions $\chi$ in \Ref{chi}. }

\noindent {\bf A4:} {\it Replace the nodal Fourier space regions
  $\Lambda^*_{r,s=\pm}$ by the sets $\Lambda^*_{s=\pm}$ in \Ref{Lrs},
  i.e., drop the restriction on $k_{s}$.} 

\noindent {\bf A5:} {\it Replace in the interaction part
of the Hamiltonian the normal ordered nodal fermion densities
\begin{equation}
  \hat{J}_{r,s=\pm}(\vp) = \left( \tPiL \right)^2
  \sum_{\vk\in\Lambda^*_s} :\!
  \hat\psi^\dag_{r,s}(\vk-\vp)\hat\psi^{\nd}_{r,s}(\vk)\!: 
\end{equation} 
by the ones in \Ref{Jrs}, i.e.\ add the umklapp terms $n\neq 0$
parallel to the Fermi surface arc.}

We thus obtain the model in \Ref{HL}--\Ref{mua}.

Note that in approximations~A4 and A5 we add terms to the nodal
fermion densities, and we therefore increase the number of interaction
terms in the Hamiltonian. Approximation~A2 partly compensates for
this. To be more specific: before A3, the number of terms included in
the nodal fermion density $J_{r,s=\pm}(\vp)$ decreases linearly with
$|p_+|$ and $|p_-|$ until $|p_\pm|=2\pi/\tilde{a}$ when it becomes
zero. However, after A4 and A5 this number of terms becomes infinite
and independent of $\vp$. It therefore is natural to restrict the
interaction terms involving nodal fermions densities by imposing a
cutoff on $\vp$. We do not know a physical argument that fixes this
cutoff in a unique manner. However, we believe that the properties of
the model are not sensitive to changes of this cutoff. We therefore
choose here a cutoff that is particularly simple; see \cite{EL} for a
more general treatment.

Our arguments above can be generalized to lattice models with more
general band relations, and our choice for the regions $\Lambda^*_{s}$
can be generalized by introducing an additional parameter allowing to
change the widths of the nodal regions $\Lambda^*_\pm$
\cite{EL}. Moreover, it is not necessary to use the simplified band
relations $\eps_{r,0}$ and interaction vertices for the antinodal
fermions, and we only do this here for aesthetic reasons. It also is
possible to relax Hypothesis~H2 \cite{dWL1}, as already mentioned.

It would be interesting to check in more detail if the approximations
above indeed do not change the low energy properties of the
model. This can be done, in principle, using our results: one can
bosonize also the terms that we added or dropped in our
approximations, and one should be able investigate the effect of these
terms on the groundstate using renormalization group methods. This is
an interesting project for the future.

\section{Bosonization and partial exact solution} 
\label{sec6}
The arguments in this section can be made mathematically precise
\cite{EL,dWL2}.

We first outline a proof of the theorem in Section~\ref{sec3}. For
that it is useful to formally write the model in \Ref{Hn}--\Ref{Jrs}
in position space as follows (we suppress arguments $\vx$ below),
\begin{equation}
\label{Hn1} 
H_n = \int\dd^2x\, \sum_{s=\pm} :\! \Bigl( \sum_{r} r v_F\,
\psi^\dag_{r,s}(-\ii \partial_s )\psi_{r,s}
+ g_1\sum_{r}J_{r,s}J_{-r,s} 
+ g_2\sum_{r,r'}J_{r,s}J_{r',-s}
\Bigr)\!:
\end{equation}
for the nodal part ($\partial_\pm$ means $\partial/\partial x_\pm$),
and
\begin{equation} 
\label{Ha1}
H_a = \int\dd^2x :\!\Bigl( \sum_r \, \psi^\dag_{r,0} (rc_F
\partial_+\partial_- -\mu_a) \psi_{r,0} 
+ g_3
\sum_{r}J_{r,0}J_{-r,0} + g_4 \sum_{s=\pm,r,r'}
\!J_{r,0}J_{r',s} \Bigr)\!:
\end{equation}
for the antinodal including the mixed parts, with $J_{r,s}(\vx)=\,
:\!\psi^\dag_{r,s}(\vx)\psi^\nd_{r,s}(\vx)\!:$ the normal ordered
fermion densities. The proper interpretation of these formulas follows
from the precise definition of the model in Fourier space: different
fermion flavors $\psi_{r,s}$ come with different Fourier space regions
$\Lambda_s^*$, and thus the spatial variable $\vx$ in
\begin{equation} 
  \psi_{r,s}(\vx)= (\tPiL)^2
  \sum_{\vk\in\Lambda^*_{s}}\hat\psi_{r,s}(\vk)\exp(\ii\vk\cdot\vx)
\end{equation} 
lives on different spaces: for the fermions $\psi_{r,+}(\vx)$ only the
variable $x_+$ is continuous while $x_-$ lives on a 1D lattice with
lattice constant $\tilde{a}$ and length $L$, and similarly for
$\psi_{r,-}(\vx)$ with $x_+$ and $x_-$ interchanged (note that the
umklapp terms with $n\neq 0$ in \Ref{Jrs} are needed for this
interpretation to be consistent with Fourier transformation). Thus the
integrals here should be (partly) interpreted as Riemann sums
\cite{EL}. In particle physics parlance, $1/\tilde a$ is a UV cutoff
needed to give a precise mathematical meaning to the model.

Thus $\psi_{r,+}(\vx)$ in \Ref{Hn1} can be treated as a collection of
1D Dirac fermions where $x_+$ plays the role of the 1D space
coordinate and $x_-$ of a flavor index.  We therefore can use the
standard mathematical results of 1D bosonization; see e.g.\ \cite{CR}
or \cite{DS}.  These imply that the densities $J_{r,+}$ obey the
following commutator relations,
\begin{equation} 
\label{JJ1} 
[J_{r,+}(\vx),J_{r',+}(\vy)] = \delta_{r,r'} r
  \frac{1}{2\pi\ii\tilde a} \partial_+\delta^2(\vx-\vy)
\end{equation}
where $\delta^2(\vx-\vy)$ here means $\delta(x_+ -y_+)
\delta_{x_-,y_-}/\tilde a$, and
\begin{equation} 
\label{bosonize}
  r \int \dd^2 x\,
  :\!\psi^\dag_{r,+}(-\ii\partial_+)\psi^\nd_{r,+}\!:\; = \tilde a
  \int \dd^2 x\, :\!J_{r,+}^2\!: 
\end{equation}
where $\int \dd^2 x$ stands for $\int\dd x_+\sum_{x_-} \tilde a$.
These and similar formulas for $\psi_{r,-}$ imply \Ref{JJ} and
\Ref{Hnboson}. The other relations stated in part (a) of the theorem
in Section~\ref{sec3} follow from well-known corresponding results in
1D.

We now describe how to exactly diagonalize the bosonized nodal
Hamiltonian $H_n$.  For simplicity we continue our discussion in
position space and ignore UV cutoffs that do not affect the results we
mention (a more detailed solution, including all cutoffs explicitly,
can be found in \cite{EL}). The commutator relations in \Ref{JJ1}
imply that the operators
\begin{equation}
\begin{split}
\partial_\pm\Phi_\pm = \sqrt{\pi\tilde a} \left(
J_{+,\pm} + J_{-,\pm} \right) \\ \Pi_\pm =
\sqrt{\pi\tilde a} \left( -J_{+,\pm} + J_{-,\pm} \right) 
\end{split}
\end{equation}
obey the commutator relations of standard 2D boson fields:
$[\Phi_{\pm}(\vx),\Pi_{\pm}(\vy)]=\ii\delta^2(\vx-\vy)$ etc., and by
straightforward computations one finds that the nodal Hamiltonian can
be expressed in terms of these bosons as follows,
\begin{equation} 
\label{bosons} 
H_n = \frac{v_F}2\int\dd^2 x :\!\Bigl( \sum_{s=\pm}[ (1-\gamma)
  \Pi_s^2 + (1+\gamma)
(\partial_s\Phi_s)^2] +
  2\gamma  (\partial_+\Phi_+)(\partial_-\Phi_-) \Bigr)\! :
\end{equation}
where $\gamma= V\sin(Q)/(4\pi t)$. It is important to note that this
Hamiltonian is positive definite only if $\gamma<1$, and this implies
an upper bound on the interaction strength mentioned in the theorem in
Section~\ref{sec3}, similarly as in 1D.  Fortunately, this bound is
satisfied for interesting parameter values ($V/t$ from $0$ up to $12$
or so).

We thus obtain an exact representation of the 2D analogue of
the Luttinger model as a system of non-interacting bosons
coupled linearly to the interacting antinodal fermions. The boson
Hamiltonian in \Ref{bosons} can be diagonalized by standard methods,
and we find
\begin{equation}
H_n = \cE_n+ \left(\tPiL\right)^2 \sum_{\vp} \sum_{s=\pm}
\omega_s(\vp) b^\dag_s(\vp)b^\nd_s(\vp)
\end{equation}
with some computable constant $\cE_n$, the dispersion relations
\begin{equation} 
  \label{omegapm}
  \omega_\pm(\vp) = \frac{v_F}{2\sqrt2}\sqrt{1-\gamma^2} 
  \sqrt{|\vp|^2\pm
    \sqrt{|\vp|^4-(1-{[}\gamma/(1+\gamma){]}^2)(2p_+p_-)^2 } },  
\end{equation}
and standard boson operators $b_s(\vp)$, i.e\
$[b^\nd_\pm(\vp),b_\pm^\dag(\vp')]=\delta^2(\vp-\vp')$ etc.  Note that
the anisotropy of the non-interacting systems is reduced by
$\gamma\neq 0$. Using that it is straightforward to compute the exact
energy eigenstates, energy eigenvalues, and partition function for the
model defined by the Hamiltonian in \Ref{Hn}.

One can also exactly integrate out the bosons and thus derive an
effective action for the antinodal fermions.  We found that, in a good
approximation, this action can be represented by an effective
Hamiltonian that has the same form as $H_a$ in \Ref{Ha} but with the
coupling constants changed to \cite{EL}
\begin{equation}
\label{g14}
  g_3= 2Va^2\left(1 -\frac{V}{2\sin(Q)[2\pi t+V\sin(Q)]}\right), \quad g_4=0. 
\end{equation}
It is possible to compute the CDW gap $\Delta\propto \langle
\psi^\dag_{+,0}\psi^\nd_{-,0}\rangle$ for this latter effective
antinodal model using mean field theory \cite{dWL1}. We found that
$\langle :\!\!N_a\!\!:\rangle =\langle:\!\!N_n\!\!:\rangle =0$ and
$\Delta>0$ are indeed independent of the parameter $\mu_a$ for
intermediate values of $V/t$ and in some $Q$-interval around
$Q=\pi/2$ \cite{dWL1}, in agreement with Hypothesis~H2 in
Section~\ref{sec4} .

We believe that, after bosonization, it is possible to make
mathematical sense of the continuum limit $\tilde{a}\to 0$ of the
nodal Hamiltonian, but we expect that this requires non-trivial
additive and multiplicative renormalizations, similarly as in 1D
\cite{GLR}. This would give further support to our conjecture that the
approximations in Section~\ref{sec5} do not affect the low energy
properties of the model.

\section{Final remarks} 
\label{sec7}
We argued that there exists a parameter regime away from half filling
where the antinodal fermions $\psi_{r,0}$ are gapped. We propose that,
in this regime, the nodal Hamiltonian in \Ref{Hn} alone accounts for
the low energy physics of the 2D $t$-$V$ model. This nodal model is
exactly solvable by bosonization and, in particular, it is possible to
compute all its Green's function by analytical methods. It would be
interesting to do this and thus confront this model with experimental
results on 2D correlated fermion systems. We hope to come back to this
in the near future \cite{dWL2}.

While the 2D analogue of the Luttinger model becomes particular simple
in the regime where the antinodal fermions are gapped, it can be used
also for other parameter values \cite{dWL1}. We believe that a mean
field treatment of the antinodal fermions is appropriate to compute
phase diagrams.  However, in parameter regions where mean field theory
predicts that the antinodal fermions are gapless, a treatment of the
antinodal Hamiltonian in \Ref{Ha} beyond mean field theory would be
desirable.

Our approach can be straightforwardly generalized to 2D lattice
fermion models with more complicated band relations \cite{EL,dWL1} and
spin \cite{dWL2}, similarly as in 1D \cite{Tsvelik}.

It is known that the truncated nodal model in \Ref{Hn} with $g_2=0$
has ``Luttinger liquid'' behavior after \cite{VM} but not before
Approximation~A5 \cite{ZYD}. However, we believe that our model with
$g_2>0$ is less sensitive to Approximation~A5 than the truncated model
with $g_2=0$. The reason is that the boson propagator has better
decaying properties in Fourier space for $g_2>0$ than for $g_2=0$
(since the boson dispersion relations in \Ref{omegapm} are quite
isotropic, whereas for $g_2=0$ they are $\omega_\pm(\vp)\propto
|p_\pm|$ independent of $p_{\mp}$).

Although we suggest that the model in \Ref{HL}--\Ref{Jrs} is a 2D
analogue of the Luttinger model, we emphasize that we do {\em not}
state that this model has ``Luttinger-liquid'' behavior
\cite{Anderson}: if or not this is the case is a delicate question and
remains to be seen. We plan to come back to this in the near future
\cite{dWL2}.

We finally discuss previous work related to ours. The pioneer in
higher dimensional bosonization is Luther \cite{Luther1}. To our
knowledge it was Mattis \cite{Mattis} who first made the important
observation that a 2D model similar to our nodal fermion model (i.e.\
no antinodal fermions) is exactly solvable by bosonization; see also
Ref.\ \cite{Rice,Hlubina} for previous work on Mattis' model. Our work
can be regarded as a particular implementation of the idea promoted by
Anderson \cite{Anderson} that 2D interacting fermions systems can be
understood by bosonization in each direction of some Fermi surface. An
earlier implementation of this idea and starting with a square Fermi
surface is by Luther \cite{Luther2}, but different from him we
bosonize chains in position space (rather than Fourier space) and thus
can avoid certain approximations in the treatment of the interactions.
Various other implementations of the idea to bosonize 2D fermion
systems appeared in the literature before but seem to differ in detail
from ours; see e.g.\ \cite{review,P} and references therein.  We
finally mention other work that was important as inspiration for us,
namely renormalization group studies of the 2D Hubbard model
\cite{FRS,HSFR} and work by Schulz \cite{Schulz} emphasizing the
significance of the antinodal fermions for 2D lattice fermion systems.

\bigskip

\noindent {\bf Acknowledgments.}  
I acknowledge helpful discussions with Alexios Polychronakos, Asle
Sudb{\o}, Manfred Salmhofer and Mats Wallin. I also would like to
thank anonymous referees, Vieri Mastropietro, and in particular Jonas
de Woul for constructive criticism that helped me to improve this
paper.  This work was supported by the Swedish Science Research
Council~(VR), the G\"oran Gustafsson Foundation, and the European
Union through the FP6 Marie Curie RTN {\em ENIGMA} (Contract number
MRTN-CT-2004-5652).


\begin{thebibliography}{99}

\bibitem{Lutt} J. M. Luttinger: An exactly soluble model of a
  many-fermion system, J. Math. Phys. \textbf{4}, 1154 (1963)
   
\bibitem{LM} D. C. Mattis and E. H. Lieb: Exact solution of a
  many-fermion system and its associated boson field,
  J. Math. Phys. \textbf{6}, 304 (1965)

\bibitem{T} S. Tomonaga: Remarks on Bloch's method of sound waves
  applied to many-fermion problems, Prog. Theor. Phys. \textbf{5}, 544
  (1950)

\bibitem{Th} W. Thirring: A soluble relativistic field theory,
  Ann. Phys. \textbf{3}, 91 (1958)

\bibitem{J} K. Johnson: Solution of the equations for the Green
  functions of a two dimensional relativistic field theory, Nuovo
  Cim. \textbf{20}, 773 (1961)

\bibitem{LLL} R. Heidenreich, R. Seiler, D. A. Uhlenbrock: The
  Luttinger model, J. Stat.  Phys. \textbf{22}, 27 (1980)

\bibitem{Haldane} F. D. M. Haldane: ''Luttinger liquid theory'' of
  one-dimensional quantum fluids: I. Properties of the Luttinger model
  and their extension to the general 1D interacting spinless Fermi
  gas, J. Phys. C \textbf{14}, 2585 (1981)
  
\bibitem{Tsvelik} A. O. Gogolin, A. A. Nersesyan, and A. M. Tsvelik:
  Bosonization and strongly correlated systems. Cambridge University
  Press, Cambridge (1998)

\bibitem{CR} See e.g.: A.L. Carey and S.N.M. Ruijsenaars: On fermion
  gauge groups, current algebras and Kac-Moody algebras, Acta
  Appl. Mat. \textbf{10}, 1 (1987)

\bibitem{DS} J. von Delft and H. Schoeller: Bosonization for beginners
  - refermionization for experts, Ann. Phys. (Leipzig) \textbf{7}, 225
  (1998)

\bibitem{Luther1} A. Luther: Tomonaga fermions and the Dirac equation
  in three dimensions, Phys. Rev. B \textbf{19}, 320 (1979)

\bibitem{Mattis} D. C. Mattis: Implications of infrared instability in
  a two-dimensional electron gas, Phys. Rev. B \textbf{36}, 745 (1987)

\bibitem{Rice} D. V. Khveshchenko, R. Hlubina and T. M. Rice:
  Non-Fermi-liquid behavior in two dimensions due to long-ranged
  current-current interactions, Phys.  Rev. B \textbf{48}, 10766
  (1993)

\bibitem{Hlubina} R. Hlubina: Luttinger liquid in a solvable
  two-dimensional model, Phys. Rev. B \textbf{50}, 8252 (1994)

\bibitem{Anderson} P. W. Anderson: ``Luttinger-liquid'' behavior of
  the normal metallic state of the 2D Hubbard model,
  Phys. Rev. Lett. \textbf{64}, 1839 (1990)

\bibitem{Luther2} A. Luther: Interacting electrons on a square Fermi
  surface, Phys. Rev. B \textbf{50}, 11446 (1994)

\bibitem{review} A. Houghton, H.-J. Kwon,
  J. B. Marston: Multidimensional bosonization,
  Adv. Phys. \textbf{49}, 141 (2000) [{\tt cond-mat/9810388}]

\bibitem{P} A. P. Polychronakos: Bosonization in higher dimensions via
  noncommutative field theory, Phys. Rev. Lett. \textbf{96}, 186401
  (2006)
  
\bibitem{EL} E. Langmann: A 2D Luttinger model, {\tt arXiv:0903.0055v3
    [math-ph]}

\bibitem{dWL1} J. de Woul and E. Langmann: Partially gapped fermions
  in 2D, J. Stat. Phys. (to appear) {\tt arXiv:0907.1277v2 [math-ph]}

\bibitem{dWL2} J. de Woul and E. Langmann (work in progress) 

\bibitem{Shankar} R. Shankar: Renormalization-group approach to
  interacting fermions, Rev. Mod. Phys. \textbf{66}, 129 (1994)

\bibitem{ARPES} For review see e.g.\ A. Damescelli, Z. Hussain and
  Z.-X. Shen: Angle-resolved photoemission studies of the cuprate
  superconductors, Rev. Mod. Phys. \textbf{75}, 473 (2003)

\bibitem{FRS} N. Furukawa, T. M. Rice, and M. Salmhofer: Truncation of
  a two-dimensional Fermi surface due to quasiparticle gap formation
  at the saddle points, Phys. Rev. Lett. \textbf{81}, 3195 (1998)

\bibitem{HSFR} C. Honerkamp, M. Salmhofer, N. Furukawa, and
  T. M. Rice: Breakdown of the Landau-Fermi liquid in two dimensions
  due to umklapp scattering, Phys. Rev. B \textbf{63}, 035109 (2001)

\bibitem{Schulz} H. J. Schulz: Fermi-surface instabilities of a
  generalized two-dimensional Hubbard model, Phys. Rev. \textbf{39},
  2940 (1989)

\bibitem{LW} E. Langmann and M. Wallin: Mean field magnetic phase
  diagrams for the two dimensional $t$-$t'$-$U$ Hubbard model,
  J. Stat. Phys. \textbf{127}, 825 (2007)


\bibitem{GLR} H. Grosse, E. Langmann and E. Raschhofer: On the
  Luttinger-Schwinger model, Annals of Phys. (N.Y.) \textbf{253}, 310
  (1997)

\bibitem{VM} V. Mastropietro: Luttinger liquid fixed point for a
  two-dimensional flat Fermi surface, Phys. Rev. B \textbf{77}, 195106
  (2008)

\bibitem{ZYD} A. T. Zheleznyak, V. M. Yakovenko and
  I. E. Dzyaloshinskii: Parquet solution for a flat Fermi surface,
  Phys. Rev. B \textbf{55}, 3200 (1997)


\end{thebibliography}
\end{document}